# Simulations for Augmented Reality Evaluation of Tools for Mass Casualty Incident Triage

Cassidy R. Nelson[1]  Joseph L. Gabbard[2]  Jason B. Moats[3]  Ranjana K. Mehta[4]
University of Utah*  Virginia Tech  Texas A&M  UW Madison

**ABSTRACT**

Mass casualty incidents (MCIs) are a high-risk, sensitive domain with profound implications for patient and responder safety. Augmented reality has shown promise as an assistive tool for high-stress work domains and MCI triage both in the field and for pre-field training. However, the vulnerability of MCIs makes it challenging to evaluate new tools designed to enhance MCI response. In other words, profound evolutions like the integration of augmented reality into field response require thorough proof-of-concept evaluations before being launched into real-world response. This paper describes two progressive simulation strategies for augmented reality that bridge the gap between computer-based simulation and actual field response.

**Index terms**: Mass casualty incident, triage, augmented reality, simulation

## 1 INTRODUCTION

Mass casualty incidents (MCIs) overwhelm available resources with deadly consequences due to "… poor leadership, planning, communication, and resource management" [1-4]. Triage assesses and prioritizes patient injuries to determine who gets what resources, faster care, and priority evacuation [5]. Augmented reality has shown promise in facilitating complex and vulnerable domains, including emergency response [7]. However, feasibility testing of such tools requires sensitive and thorough evaluation. E.g., lab-studies alone are not sufficient to guarantee safety during actual chaotic MCI response. Thus, we propose two progressive simulation strategies that could be used for future AR tool testing. Section 2.3 describes a virtual patient simulation, and 2.4 outlines a 'real' patient simulation with patient actors. The proposed simulations leverage a Hololens 2 due to its extensive industrial applications broadly and in MCIs already [8], [9].

### 1.1 Contributions of this Work

This work offers a first-of-its-kind intensive-simulation based evaluation for field triage. This work will outline a 'virtual' patient simulation and a 'real' (patient actor) simulation. Given the high cost of patient-actor simulations, there is a growing prominence of AR/VR simulations for triage training [10], [11], and these simulations have further been designed to facilitate training [12], [13]. Moreover, we posit that such simulations could be used to evaluate other potential AR support strategies. Future work will discuss how these simulations were used to evaluate the Augmented Reality Triage Tool Suite [8], [9].

## 2 METHODS & DISCUSSION

The current gold standard for MCI simulation training is to send responders to one of a select few national MCI training facilities in the country. We partnered with Disaster City at Texas A&M, who specializes in MCI triage for emergency responders all over the country, to create and execute the virtual patient and patient actor simulations. The simulations were developed by three human factors engineers and a primary subject matter expert (SME) with over 30 years as an EMT and as an MCI trainer for other emergency medical service (EMS) providers.

### 2.1 SALT Triage Process

To understand the simulations, we must first describe the gold standard MCI Triage Process: the SALT (the Sort, Assess, Lifesaving Intervention, Treatment/Transport) Framework [14], [15]. SALT is a 'wave' based flow chart with ≥2 waves. Wave one is to roughly sort patients based on their ability to walk and respond to commands appropriately. Patients who can walk and understand commands are assessed last, patients who can understand but cannot walk second, and patients who cannot do either are assessed first. Wave 2 is where patients are actually

[1]email: cassidy.nelson@utah.edu *(formerly cassidynelson@ohio.edu)
[2]email: jgabbard@vt.edu
[3]email: jbmoats@tamu.edu
[4]email: rmehta38@wisc.edu

assigned one of five possible triage categories; **Black**: deceased patients (last priority); **Grey**: patients expected to expire as current lifesaving resources are not likely to save (third priority); **Red**: patients needing immediate care in critical condition that can potentially be saved (first priority); **Yellow:** Patients who can withstand delayed care after receiving stabilizing care in the field but will still need follow-up (second priority), and **Green:** patients experiencing minimal to no injuries (fourth priority).

## 2.2 Defining Simulation Parameters

Our primary SME generated a "Master Case List" of patient data for 20 patients that appropriately fall within specific SALT triage categories so the Virtual and 'Real' simulations had patient parameters to triage against. Given the relative ease of assessing a green/minimal or black/dead category and the fuzziness between other categories, we used variable category weightings to offer more opportunity to parse these 'fuzzy' delineations (like between grey/red). Our case list included three black/dead patients, four grey/expectant, five red/immediate, five yellow/delayed, and three green/minimal as defined by the SALT framework [15]. Both simulation applications have an 'author mode' that can be toggled by an experimenter or trainer by saying 'author mode'. This mode allows the user to place where a virtual patient should appear in the world via locking to the world spatial mesh passively captured by Hololens 2. Once done, the facilitator simply says 'author mode' again to see the 'trainee view' of the virtual patients. The app connects to a local server where facilitators can tweak parameters and where real-time data is sent. This data includes tracked hand data, gaze, interaction data (what was clicked at what time), time spent within a triage task, and triage accuracy.

## 2.3 Virtual Patient Simulation

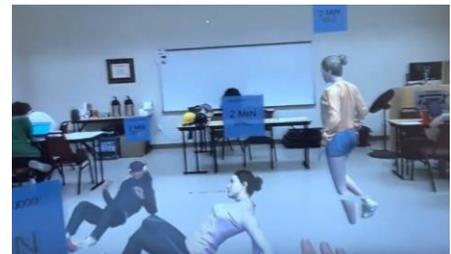

Figure 1: Participant view of Virtual Patients

Due to the probable space limitations of future at-home training and testing simulations, the virtual simulation task is smaller than the 'real' task. This task requires triaging five patients in ten minutes. Each simulation loads in 5 patients randomly assigned attributes based on that master case list with the following two constraints: 1) There needs to be one white woman, one black woman, one black man, one white man, and one random (to ensure representative study stimuli [16]), and 2) there is one patient from each triage category represented. Responders can interact with virtual patients to capture blood pressure by putting their palm on the bicep of the virtual patient to get a read-out, pulse by putting two fingers on the patient's wrist to generate an auditory heartbeat, respiration rate by putting their head next to the virtual patient's chest to listen to and count breaths, and finally cognitive status by asking "can you wave" or "show me where it hurts" where the virtual patient would then wave or gesture toward their injury. Virtual patients have variable animations (see figure 1) captured via mo-cap to communicate injury status. Moreover, some virtual patients have 'moulage', or blood and other injuries visible.

## 2.4 'Real' Patient Actor Simulation

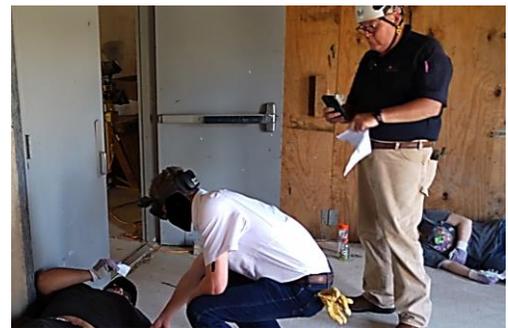

Figure 2: Participant, facilitator, & patient actor

For emergency medical units with enough financial wherewithal to send their responders to a National Training Facility like Disaster City, we developed another simulation that is closer to live field response. We want to note, however, that it is possible that future response units with AR headsets could adopt this simulation strategy within their own stations if provided with the master case list and the simulation application.

The simulation setting is of a fertilizer explosion. We were able to recruit twenty patient actors via a local acting troupe that were racially and gender diverse. The same aforementioned master case list defines the parameters of injuries for the patient actors. Moreover, each actor is given a script to follow that assigns a movement (holding their arm, rocking, etc.) and provides the cognitive orientation voice lines, and actors are then given a specific location to stand/sit/lie. We encourage future researchers to leverage author mode for the 'real' simulation as well to ensure the simulation can be kept consistent while mixing up triage categories and certain severities are not always clumped together. Using 'author mode,' virtual patients can be placed in the physical world and then turned invisible to the user. This means that physical actors can stand on the same place and ensure that data is logged for a specific location and patient ID, and cross checked for accuracy in real time. The 'real' patient actor condition task is built to triage twenty patients in ten minutes. This is a deliberately larger set from the virtual patient simulation for two reasons. 1) the real simulation presupposes a larger evaluation space like a field house, and 2) MCI triage is a highly collaborative effort. To afford team evaluations of shared AR collaboration, this task is designed as a 'team task' with two responders and thus demands more patients to triage. Finally, as we cannot simulate a healthy actor having a dangerously low heart rate, etc., each participant responder is followed by a training facilitator at Disaster City (see Figure 2). When the participant goes to capture something like heart rate, the facilitator will tell them what the value should be. Patients are scattered in the 'theatre', a deliberately derelict building with trip hazards and scattered debris for training purposes.

## 3  CONCLUSION

This paper describes two progressive AR simulations that depart from standard desktop evaluation and training strategies to more closely approximate real-field response. These simulations were designed in partnership with Disaster City at TEEX. Such simulations will be critical to feasibility testing of AR tools before they are adopted in the field. Future work will use these simulations to measure potential AR applications for MCI triage while simultaneously assessing simulation efficacy.

### 3.1  Acknowledgements

This material is based upon work supported by the National Science Foundation under Grant No. 2033592.

**REFERENCES**
[1] World Health Organization, "Mass Casualty Management Systems Strategies and guidelines," 2007.
[2] A. K. Donahue and R. V. Tuohy, "'Lessons we don't learn: a study of the lessons of disasters; why we repeat them; and how we can learn them," *Homeland Security Affairds*, vol. 2, no. 2, pp. 1–28, 2006.
[3] M. Sommer, G. S. Braut, and O. Nja, "A model for learning in emergency response work," *International Journal of Emergency Management*, 2013, doi: 10.1504/IJEM.2013.055161.
[4] U. Rosenthal, R. A. Boin, and L. K. Comfort, *Managing Crises: Threats, Dilemmas, Opportunities,*. Springfield, Illinois, 2001.
[5] M. D. R. F. Fernández, C. I. T. Bernabe, and R. R. Rodríguez, "Red cross triage app design for augmented reality glasses," *ACM International Conference Proceeding Series*, vol. 03-05-Nove, no. 2901, pp. 11–14, 2014, doi: 10.1145/2676690.2676697.
[6] P. M. Furbee *et al.*, "Realities of rural emergency medical services disaster preparedness," *Prehosp Disaster Med*, vol. 21, no. 2, pp. 64–70, 2006, doi: 10.1017/S1049023X0000337X.
[7] E. Argo, T. Ahmed, S. Gable, C. Hampton, J. Grandi, and R. Kopper, "Augmented Reality User Interfaces for First Responders: A Scoping Literature Review," Available: http://arxiv.org/abs/2506.09236
[8] C. R. Nelson, J. L. Gabbard, J. B. Moats, and R. K. Mehta, "User-Centered Design and Evaluation of ARTTS: an Augmented Reality Triage Tool Suite for Mass Casualty Incidents," in *Proceedings - 2022 IEEE International Symposium on Mixed and Augmented Reality, ISMAR 2022*, Institute of Electrical and Electronics Engineers Inc., 2022, pp. 336–345. doi: 10.1109/ISMAR55827.2022.00049.
[9] C. Rae Nelson *et al.*, "Exploring Augmented Reality Triage Tools to Support Mass Casualty Incidents," *Proceedings of the Human Factors and Ergonomics Society Annual Meeting*, vol. 66, no. 1, pp. 1664–1666, Sep. 2022, doi: 10.1177/1071181322661337.
[10] S. Khanal, U. S. Medasetti, M. Mashal, B. Savage, and R. Khadka, "Virtual and Augmented Reality in the Disaster Management Technology: A Literature Review of the Past 11 years," Apr. 11, 2022, *Frontiers Media S.A.* doi: 10.3389/frvir.2022.843195.
[11] M. del Carmen Cardós-Alonso *et al.*, "Extended reality training for mass casualty incidents: a systematic review on effectiveness and experience of medical first responders," Dec. 01, 2024, *BioMed Central Ltd*. doi: 10.1186/s12245-024-00685-3.
[12] R. R. Mohanty *et al.*, "From Discovery to Design and Implementation: A Guide on Integrating Immersive Technologies in Public Safety Training," *IEEE Transactions on Learning Technologies*, vol. 18, pp. 387–401, 2025, doi: 10.1109/TLT.2025.3555649.
[13] R. R. Mohanty *et al.*, "The Design and Evaluation of an AR-based Adaptive Triage Training for Emergency Responders," *Proceedings of the Human Factors and Ergonomics Society Annual Meeting*, Sep. 2024, doi: 10.1177/10711813241265339.
[14] E. B. Lerner *et al.*, "Mass casualty triage: An evaluation of the science and refinement of a national guideline," *Disaster Med Public Health Prep*, vol. 5, no. 2, pp. 129–137, 2011, doi: 10.1001/dmp.2011.39.
[15] Silvestri Salvatore *et al.*, "Comparison of START and SALT triage methodologies to reference standard definitions to a field mass casualty simulation," *Am J Disaster Med*, vol. 12, no. 1, pp. 27–33, 2017.
[16] T. D. Do, S. Zelenty, M. Gonzalez-Franco, and R. P. McMahan, "VALID: a perceptually validated Virtual Avatar Library for Inclusion and Diversity," *Front Virtual Real*, vol. 4, 2023, doi: 10.3389/frvir.2023.1248915.